\documentclass{elsart}
\usepackage{graphicx}

\begin{document}
\begin{frontmatter}
\title{Quantitative photoluminescence of broad band absorbing melanins: A procedure to correct for inner filter and re-absorption effects}

\author[BandLaser]{Jennifer Riesz\thanksref{correspondance}}
\author[CM]{Joel Gilmore}
\author[BandLaser]{Paul Meredith}

\thanks[correspondance]{Corresponding author. Tel.: +61 7 3365 3406; Fax: +61 7 3365 1242
E-mail address: riesz@physics.uq.edu.au}

\address[BandLaser]{Department of Physics, Soft Condensed Matter Physics Group and Centre for Biophotonics and Laser Science, University of Queensland, St. Lucia, Brisbane, Queensland 4072, Australia}
\address[CM]{Department of Physics, Condensed Matter Theory Group, University of Queensland, Brisbane, Queensland 4072, Australia}

\begin{abstract}
We report methods for correcting the photoluminescence emission and excitation spectra of highly absorbing samples for re-absorption and inner filter effects.  We derive the general form of the correction, and investigate various methods for determining the parameters.  Additionally, the correction methods are tested with highly absorbing fluorescein and melanin (broadband absorption) solutions; the expected linear relationships between absorption and emission are recovered upon application of the correction, indicating that the methods are valid.  These procedures allow accurate quantitative analysis of the emission of low quantum yield samples (such as melanin) at concentrations where absorption is significant.
\end{abstract}


\end{frontmatter}


\section{Introduction}
\begin{figure}
	\begin{center}
		\includegraphics{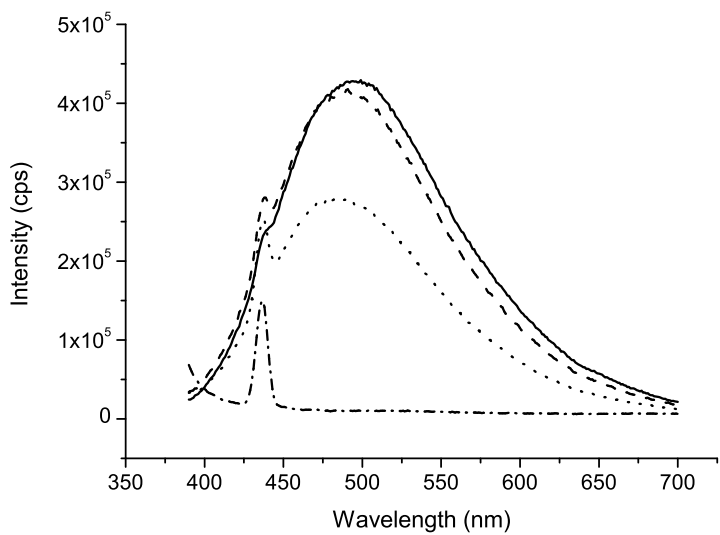}
	\end{center}
	\caption{Raw PL emission spectra (pumped at 380nm) for three synthetic eumelanin solutions: $0.005\%$ (solid), $0.0025\%$ (dashed) and $0.001\%$ (dotted) by weight concentration, and solvent background (dot-dash).}
	\label{fig:Raw}
\end{figure}

\begin{figure}
	\begin{center}
		\includegraphics{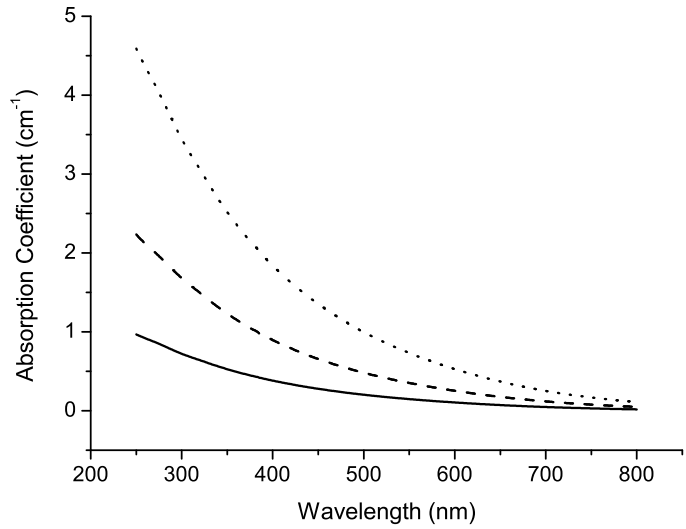}
	\end{center}
	\caption{Absorption spectra of synthetic eumelanin solutions at three concentrations: $0.005\%$ (dotted), $0.0025\%$ (dashed) and $0.001\%$ (solid) by weight concentration.}
	\label{fig:absplot}
\end{figure}

\begin{figure}
	\begin{center}
		\includegraphics{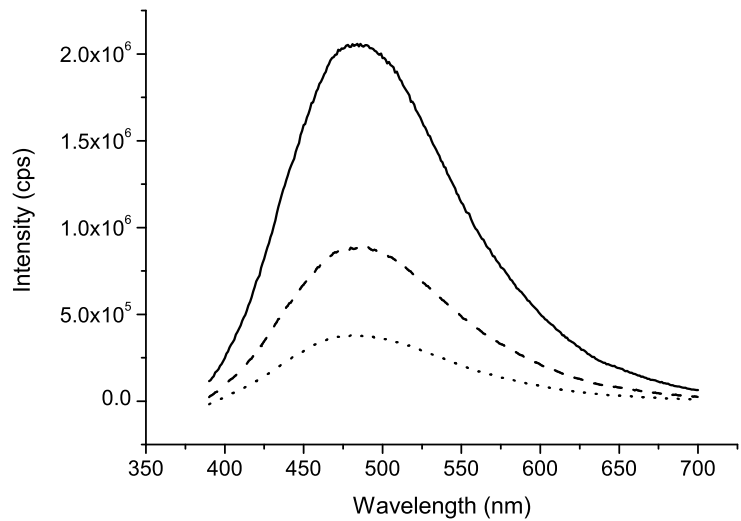}
	\end{center}
	\caption{PL emission spectra corrected using Method 1 (pumped at 380 nm) for three synthetic eumelanin solutions: 0.005$\%$ (solid), 0.0025$\%$ (dashed) and 0.001$\%$ (dotted) by weight concentration.  Errors were calculated according to equation \ref{eq:generaluncert}.}
	\label{fig:corrected}
\end{figure}

Melanins are an important class of dark biological pigment known to be responsible for photoprotection of the skin, hair and eyes of many different species \cite{Prota92,Vitkin94,Kollias91}.  Of the two types found in humans (eumelanin: black, and pheomelanin: red-brown) eumelanin is the most common \cite{Teuchner00}.  Paradoxically, it is also believed that melanins are directly involved in the UVA-induced photochemical processes that lead to DNA damage in skin cells \cite{Hill95,Nofsinger99,Menon77}.  The biological structure-property-function relationships of melanins and their connections to melanoma formation have become the foci of intense scientific interest and extensive study.  Despite this activity, there remain many unanswered questions regarding its structure and properties \cite{Stark03,Forest98,McGuiness74}.  This fact is in part due to melanin's strong, broadband and highly non-linear absorbance (Figure \ref{fig:absplot}), very low quantum yield \cite{Meredith04} and insolubility \cite{Prota92,Mosca99} which make spectroscopic analysis challenging.  Emission levels are so low that they must be measured at concentrations where re-absorption and inner filter effects (attenuation of the probe beam) strongly distort the resulting spectra (Figures \ref{fig:Raw} and \ref{fig:corrected}).  It is clear that a procedure to accurately and easily correct for these effects is required.  This would allow future spectroscopic measurements of eumelanin (and other strongly absorbing species) to be analysed accurately and quantitatively.  Re-absorption correction methods have been previously applied to melanin \cite{Gallas87}; the work reported here is a far more detailed analysis of the assumptions implicit in these methods, and of the effectiveness of the procedures.

\section{Calculations}
\subsection{General Form of the Correction}
\begin{figure}
	\begin{center}
		\includegraphics[width=10cm]{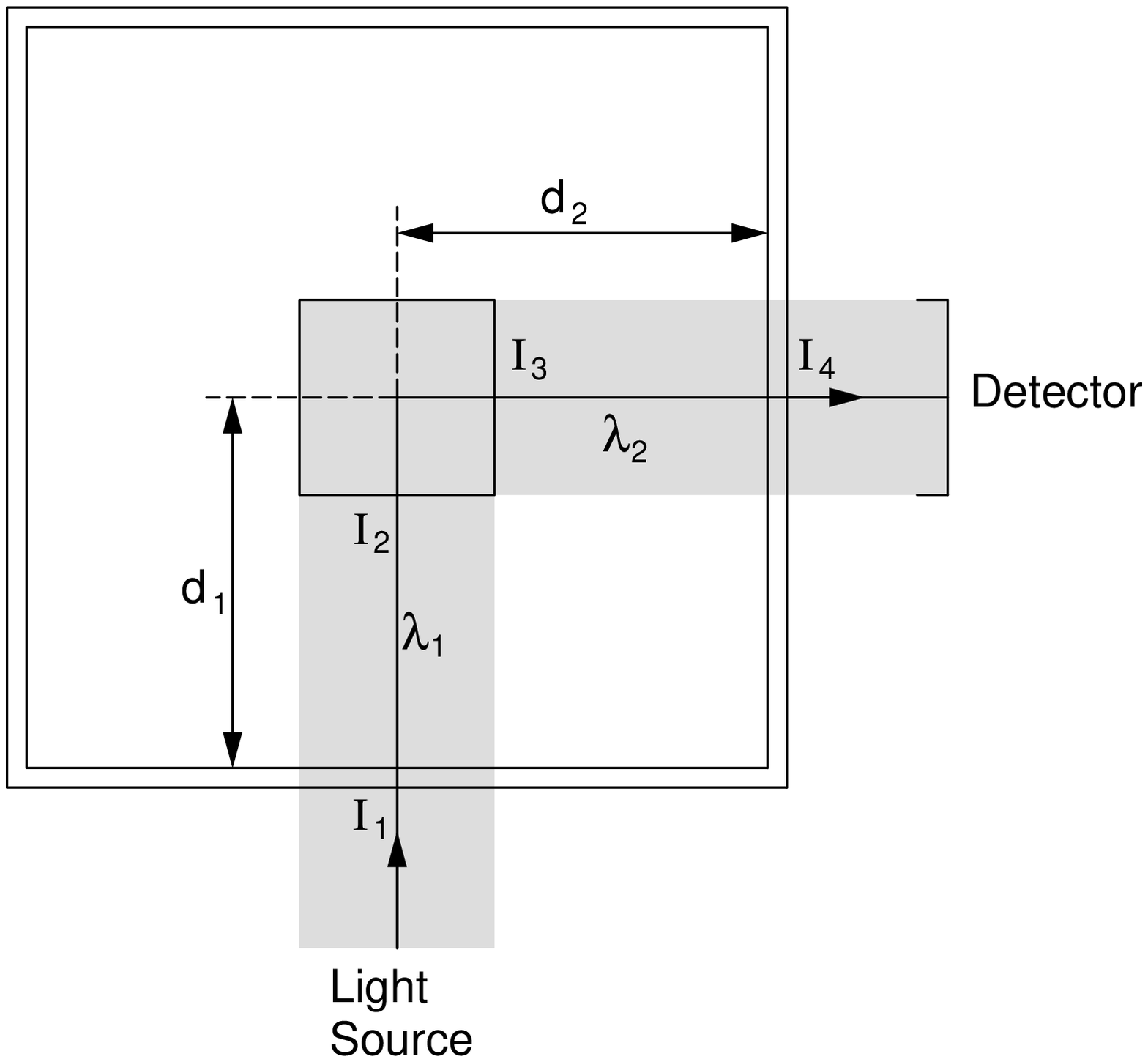}
	\end{center}
	\caption{Geometry of the PL and PLE cuvette (view from above).  Light at wavelength $\lambda_1$ travels a distance $d_1$ into the cuvette, where it excites photoluminescence (at a wavelength $\lambda_2$).  This emission then travels a distance $d_2$ through the solution to the detector.  $I_1$ is the initial intensity of excitation light, which is attenuated to intensity $I_2$ at the centre of the cuvette (inner filter effect).  $I_3$ is the initial intensity of the emission from the excitation volume, which is attenuated to intensity $I_4$ when it reaches the detector (re-absorption).}
	\label{fig:cuvette}
\end{figure}

The following derivation is for a collimated excitation beam incident upon a square cross-section cuvette (containing the solution of interest), with collection at $90^\circ$ with respect to excitation (and in the same horizontal plane), as shown in Figure \ref{fig:cuvette}. This is the most common geometry used, although this correction method could easily be applied to other geometries.  We assume a small excitation volume in the cuvette (defined by the emission and excitation slit widths) so that we can apply the Beer Lambert Law \cite{Lakowicz99} directly.  From Figure \ref{fig:cuvette}, for the incoming and outgoing light, we have:
 		 			       
\begin{eqnarray}
I_2 &=& I_1e^{-\alpha_1d_1} \label{eq:Beerin}\\
I_4 &=& I_3e^{-\alpha_2d_2} \label{eq:Beerout}
\end{eqnarray}
 		 			       
where $I_1$ to $I_4$ are the intensity of the beam at the locations indicated in Figure \ref{fig:cuvette}, $d_1$ and $d_2$ are the path lengths into and out of the cuvette, and $\alpha_1 \equiv \alpha(\lambda_1)$ and $\alpha_2 \equiv \alpha(\lambda_2)$ are the absorption coefficients along those paths (dependant upon the excitation and detection wavelengths, $\lambda_1$ and $\lambda_2$ respectively).

Let the subscript $m$ denote intensities affected by re-absorption and inner filter effects (`measured' intensities), and the subscript $c$ denote intensities corrected for these effects.  We can then split the intensities $I_3$ and $I_4$ into components as follows:

\begin{eqnarray}
I_3 &=& F_c\left(I_2,\lambda_1,\lambda_2\right)+B_c\left(I_2,\lambda_1,\lambda_2\right)+R_c\left(I_2,\lambda_1,\lambda_2\right) \\
I_4 &=& F_m\left(I_1,\lambda_1,\lambda_2\right)+B_m\left(I_1,\lambda_1,\lambda_2\right)+R_m\left(I_1,\lambda_1,\lambda_2\right) 
\end{eqnarray}

where $F$ is the component due to fluorescence of the sample (which we seek to recover), $B$ is the background fluorescence (measured from the solvent alone) and $R$ is the component due to Raman scattering from the solvent \cite{Hecht87}.  $\lambda_1$ and $\lambda_2$ are the excitation and emission wavelengths respectively.   In normal situations the fluorescence, Raman scattering and background components are all linearly dependant upon the illuminating intensity \cite{Lakowicz99,Hecht87}, so we can assume:

\begin{eqnarray}
F_c(bI) & = & bF_c(I)\\
B_c(bI) & = & bB_c(I)\\
R_c(bI) & = & bR_c(I) \label{eq:last}
\end{eqnarray}

where $b$ is an arbitrary constant.  Combining equations \ref{eq:Beerin} to \ref{eq:last} we find:

\begin{equation} \label{eq:Fullexpress}
F_c(I_1) = e^{\alpha_1d_1+\alpha_2d_2}\left[F_m(I_1)+B_m(I_1)+R_m(I_1)\right]-\left[B_c(I_1)+R_c(I_1)\right]
\end{equation}

In a more convenient form:

\begin{equation}
F_{corrected} = kF_{measured} - F_{background}
\end{equation}

where

\begin{equation} \label{eq:corfac}
k = k\left(\lambda_1, \lambda_2\right) = e^{\alpha_1d_1+\alpha_2d_2}
\end{equation}

and $F_{measured}$ (the emission we would actually measure) and $F_{background}$ (the background emission we would measure from the solvent alone) are given by the bracketed expressions in equation \ref{eq:Fullexpress}. $F_{corrected}$ is the quantity we seek - the emission due to the fluorophore alone, corrected for inner filter and re-absorption effects.  Using standard propogation of random, independant errors \cite{Taylor82}, the uncertainty in the corrected photoluminescence will then be given by:

\begin{equation} \label{eq:generaluncert}
\Delta F_{corrected} = kF_{measured}\sqrt{\alpha_1^2\Delta d_1^2+\alpha_2^2\Delta d_2^2}
\end{equation}

where we have assumed that the uncertainty in $d_1$ and $d_2$ dominates (this was found to be the case for our system).

$\alpha_1$ and $\alpha_2$ can be determined from a previously measured absorption spectrum.  For a photoluminescence (PL) measurement $\alpha_1$ will be a constant (the absorption coefficient at the excitation wavelength), whereas $\alpha_2$ will be wavelength dependant.  For a photoluminescence (PLE) measurement, $\alpha_2$ will be constant (the absorption coefficient at the detection wavelength) and $\alpha_1$ will be wavelength dependant.  The only remaining parameters, $d_1$ and $d_2$, can be determined in a variety of ways, as follows.


\subsection{Method 1: Direct Measurement}
By assuming that the small excitation volume is in the centre of the cuvette, $d_1$ and $d_2$ can simply be calculated from the known dimensions of the system (as shown in Figure \ref{fig:cuvette}).  For a convential spectroscopic system this assumption will be valid, and the geometry well known.


\subsection{Method 2: Raman Peak Attenuation}
If direct measurement of the system is not possible (in thin films for example, or another unique geometry) then we seek another measure of the re-absorption.  The attenuation of the signal associated with Raman scattering of the probe by the solvent molecules \cite{Hecht87} (the `Raman scattering peak') provides an alternative method of determining the necessary correction factor.  From equation \ref{eq:Fullexpress} we see that

\begin{equation}
k(\lambda_1,\lambda_2)R_m(I_1,\lambda_1,\lambda_2) = R_c(I_1,\lambda_1,\lambda_2)
\end{equation}

since only $R_m$ can respond to a change in $R_c$, and $R_c = 0$ must imply $R_m = 0$.  Combining with equation \ref{eq:corfac} for $k$ we find:

\begin{equation}
\left(\frac{1}{\alpha_2}\right)\ln\left(\frac{R_c}{R_m}\right) = \left(\frac{\alpha_1}{\alpha_2}\right)d_1+d_2
\end{equation}

If excitation at wavelength $\lambda_{1R}$ gives a Raman scattering peak at wavelength $\lambda_{2R}$ we have:

\begin{equation} \label{eq:Meth2}
\left(\frac{1}{\alpha(\lambda_{2R})}\right)\ln\left(\frac{R_c(\lambda_{2R})}{R_m(\lambda_{2R})}\right) = \left(\frac{\alpha(\lambda_{1R})}{\alpha(\lambda_{2R})}\right)d_1+d_2
\end{equation}

\begin{figure}
	\begin{center}
		\includegraphics{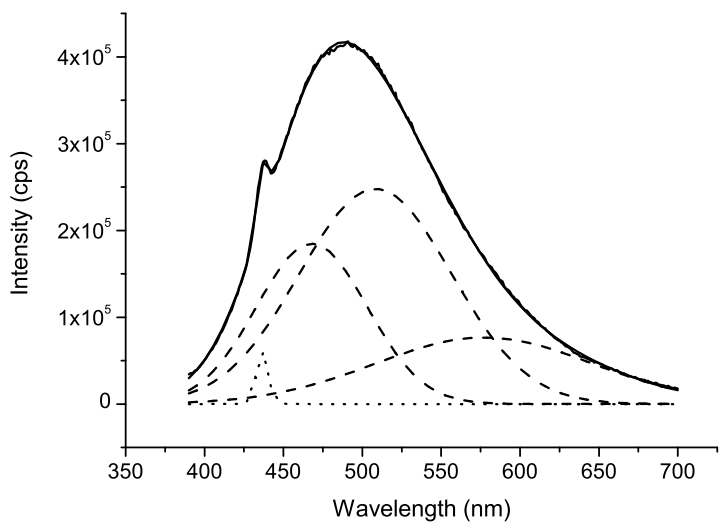}
	\end{center}
	\caption{PL (380nm excitation) of 0.0025$\%$ eumelanin solution with multiple Gaussian fitting (dashed: Gaussian components, solid: data and resultant fit, dotted: Raman scattering component fit).}
	\label{fig:RamFit}
\end{figure}

By exciting the solution at a variety of wavelengths we can plot the above function such that $d_1$ and $d_2$ are the gradient and y-intercept respectively.  $R_c$ and $R_m$ can be determined by fitting the PL spectra to multiple Gaussians, as shown in Figure \ref{fig:RamFit}.  $R_c$ and $R_m$ are the amplitudes of the Gaussian  Raman scattering peak in the background and fluorophore spectra respectively.  The uncertainty in $d_1$ and $d_2$ can be determined from the linear regression, and used in equation \ref{eq:generaluncert}.

This method theoretically allows correction of the emission spectra even if the sample geometry is unknown (such as in thin films), as long as the Raman peak is clearly visible.  Even in the well defined square cuvette geometry, this method potentially allows for the fact that the illumination volume may not be directly in the centre of the cuvette.  


\subsection{Method 3: Raman Peak Attenuation Approximation}
If it were not possible to apply Method 2 (as shall be discussed in the results section), then the following approximations may allow correction of unique system geometries.  If re-absorption dominates over inner-filter effects ($\alpha_2d_2 \gg \alpha_1d_1$) we can make the approximation:

\begin{equation} \label{eq:approx}
k=e^{\alpha_2d_{eff}}
\end{equation}

where $d_{eff}$ is an effective path length.   This may be the case either because of the geometry of the system, or because of the absorbance profile of the sample.  Following the same procedure as for Method 2 (solving for $d_{eff}$ at the excitation wavelength $\lambda_{1R}$ that gives a Raman scattering peak at wavelength $\lambda_{2R}$) we find 

\begin{equation}
d_{eff} = \left(\frac{1}{\alpha(\lambda_{2R})}\right)\ln\left(\frac{R_c(\lambda_{2R})}{R_m(\lambda_{2R})}\right)
\end{equation}					
 					
Hence the full correction is:

\begin{equation}
F_{corrected}(\lambda_2) = \exp\left(\frac{\alpha(\lambda_2)}{\alpha(\lambda_{2R})}\ln\left(\frac{R_c(\lambda_{2R})}{R_m(\lambda_{2R})}\right)\right)F_{measured}(\lambda_2)-F_{background}(\lambda_2)
\end{equation}

with uncertainty:

\begin{equation} \label{eq:uncert3PL}
\Delta F_c(\lambda_2) = \frac{\alpha(\lambda_2)k(\lambda_2)F_m(\lambda_2)}{\alpha(\lambda_{2R})}\sqrt{\left(\frac{\Delta R_c(\lambda_{2R})}{R_c(\lambda_{2R})}\right)^2+\left(\frac{\Delta R_m(\lambda_{2R})}{R_m(\lambda_{2R})}\right)^2}
\end{equation}

assuming random, independant errors, dominated by the uncertainty in $R_c$ and $R_m$ \cite{Taylor82}.  $R_c$ and $R_m$ can be determined as for Method 2.

Note that although equation \ref{eq:approx} no longer explicitly includes the excitation wavelength (as the correction factor did in Method 1), $d_{eff}$ will be dependant upon $\lambda_1$ due to the non-linearity of the absorption profile. Hence $d_{eff}$ must be recalculated for PL spectra measured at different excitation wavelengths unless the absorption coefficients at these excitation wavelengths ($\alpha_1$) are very similar, or $\alpha_2d_2$ is so much larger than $\alpha_1d_1$ that the difference is negligible.  

This also means that each point in a PLE spectrum will have a different $d_{eff}$.  Since this is impractical to calculate (unless the absorption profile is constant at all excitation wavelengths), we can instead manipulate the geometry of the system so that inner filter effects are more significant than re-absorption.  This can be achieved with a rectangular cuvette, orientated so that the excitation path length ($d_1$) is significantly longer than the emission path length ($d_2$).  We can then make the alternate approximation:

\begin{equation}
k=e^{\alpha_1d_{eff}}
\end{equation}

giving:

\begin{equation}
d_{eff} = \left(\frac{1}{\alpha(\lambda_{1R})}\right)\ln\left(\frac{R_c(\lambda_{2R})}{R_m(\lambda_{2R})}\right)
\end{equation}	

Hence the full correction for PLE spectra is:

\begin{equation}
F_{corrected}(\lambda_1) = \exp\left(\frac{\alpha(\lambda_1)}{\alpha(\lambda_{1R})}\ln\left(\frac{R_c(\lambda_{2R})}{R_m(\lambda_{2R})}\right)\right)F_{measured}(\lambda_1)-F_{background}(\lambda_1)
\end{equation}

with uncertainty:

\begin{equation} \label{eq:uncert3PLE}
\Delta F_c(\lambda_1) = \frac{\alpha(\lambda_1)k(\lambda_1)F_m(\lambda_1)}{\alpha(\lambda_{1R})}\sqrt{\left(\frac{\Delta R_c(\lambda_{2R})}{R_c(\lambda_{2R})}\right)^2+\left(\frac{\Delta R_m(\lambda_{2R})}{R_m(\lambda_{2R})}\right)^2}
\end{equation}

once again assuming random, independant errors, dominated by the uncertainty in $R_c$ and $R_m$ \cite{Taylor82}.

Note that since the Raman scattering peak is not clearly visible in PLE spectra, the effective path length ($d_{eff}$) must be calculated from the appropriate PL spectrum (with the cuvette orientated so that inner filter effects dominate).  Any excitation wavelength is sufficient to determine $d_{eff}$, as long as the Raman peak is clearly visible.

\section{Experimental}
\subsection{Sample Preparation}  
Synthetic eumelanin (dopamelanin) derived from the non-enzymatic oxidation of tyrosine was purchased from Sigma Aldrich, and used without further purification. Eumelanin solutions were prepared at a range of concentrations ($0.001\%$ to $0.005\%$) by weight macromolecule in high purity $18.2M\Omega$  MilliQ de-ionised water. To aid solubility, the pH of the solutions was adjusted using 0.01 M NaOH to $\sim11.5$, and the solutions gently heated with stirring. Under such conditions, pale brown, apparently continuous eumelanin dispersions were produced. Fluorescein was purchased from Sigma Aldrich and used without further purification to prepare standard solutions at ten different concentrations varying from $1.2\times10^{-4}\%$ to $5\times10^{-6}\%$ by weight in 0.1M NaOH solution ($18.2M\Omega$  MilliQ de-ionised water). Fluorescein and eumelanin concentrations were chosen so as to maintain absorbance levels within the range of the spectrometer.

\subsection{Absorption Spectrometry}  
Absorption spectra between 200nm and 800nm were recorded for the synthetic eumelanin and fluorescein solutions using a Perkin Elmer Lambda 40 spectrophotometer. An integration of 2nm, scan speed of 240nm/min and slit width of 3nm bandpass were used. Spectra were collected using a quartz 1cm square cuvette. Solvent scans (obtained under identical conditions) were used for background correction.

\subsection{Photoluminescence Emission Spectrometry}
Photoluminescence emission spectra for the eumelanin and fluorescein solutions were recorded for all concentrations using a Jobin Yvon FluoroMax 3 Fluorimeter. Emission scans were performed between 400nm and 700nm using an excitation wavelength of 380nm for the eumelanin samples, and 490nm for the fluorescein samples. A band pass of 3nm and an integration of 0.3s were used.   Background scans were performed under identical instrumental conditions using the relevant solvents.   Spectra were collected using a quartz 1cm square cuvette. Spectra were automatically corrected to account for differences in pump beam power at different excitation wavelengths using a reference beam.

\subsection{Photoluminescence Excitation Spectrometry}
Photoluminescence excitation spectra for the eumelanin solutions were recorded using a Jobin Yvon FluoroMax 3 Fluorimeter. Excitation scans were performed between 300nm and 465nm using a detection wavelength of 485nm.  A band pass of 3nm and an integration of 0.3s were used. Background scans were performed under identical instrumental conditions using the relevant solvents. A rectangular cuvette orientated such that $d_1 = 0.5$cm, $d_2 = 0.2$cm was used to reduce re-absorption effects in comparison to inner filter effects.  Spectra were automatically corrected to account for differences in pump beam power at different excitation wavelengths using a reference beam.


\section{Results and discussion}

\subsection{Method 1: Direct Measurement}
The uncorrected PL spectrum of eumelanin for three different concentrations is shown in Figure \ref{fig:Raw}.  Inner filter and re-absorption effects are clearly evident; the intensity does not increase linearly with concentration, the peak shifts to lower energies at higher concentrations, and the Raman scattering peak (at approximately 436nm) is increasingly attenuated as the concentration increases (such that subtraction of the background leaves `holes' in the spectra).  The shift in peak position is due to the nonlinearity of eumelanin absorbance (Figure \ref{fig:absplot}).

\begin{figure}
	\begin{center}
		\includegraphics{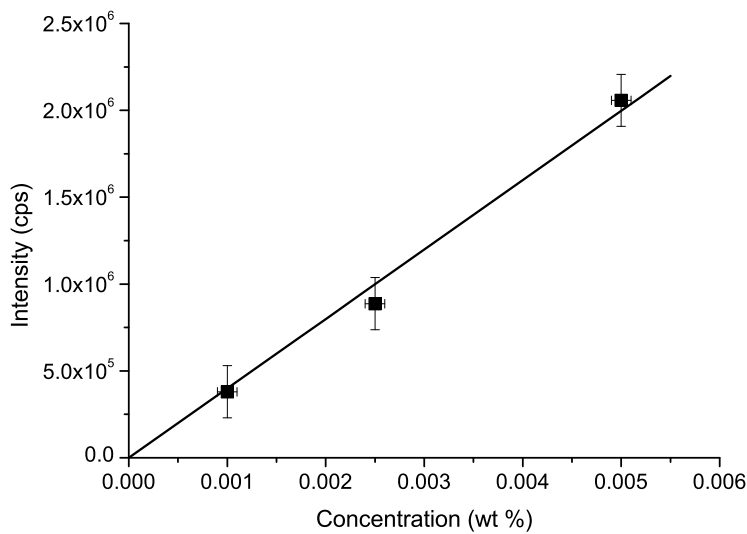}
	\end{center}
	\caption{PL emission peak intensity (corrected using Method 1) vs. concentration for three synthetic eumelanin solutions: 0.005$\%$, 0.0025$\%$ and 0.001$\%$ by weight concentration. The samples were pumped at 380 nm. Intensity errors were calculated according to equation \ref{eq:generaluncert}.}
	\label{fig:Intconc}
\end{figure}

Figure \ref{fig:corrected} shows the same PL spectra corrected using Method 1 as outlined above, with measured path lengths $d_1 = d_2 = 0.50$cm.  The peaks now align very closely (to within 2nm), the Raman peak is completely removed by subtraction of the background, and the peak intensity is now linear with concentration, as shown in Figure \ref{fig:Intconc}.
  
\begin{figure}
	\begin{center}
		\includegraphics{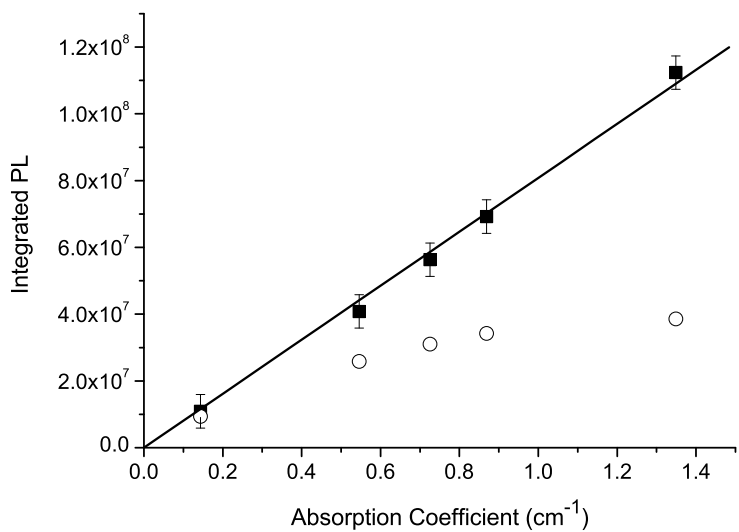}
	\end{center}
	\caption{Integrated PL emission vs. absorption coefficient at the excitation wavelength (380 nm) for five melanin solutions (0.001$\%$ to 0.005$\%$): uncorrected data (open circles) and corrected with Method 1 (solid squares with linear regression). Errors were calculated according to equation \ref{eq:generaluncert}.}
	\label{fig:MelQ380}
\end{figure}

Figure \ref{fig:MelQ380} shows the integrated PL against the absorbance at the excitation wavelength for various concentrations of eumelanin.  Open circles indicate the uncorrected data, and closed squares show the data after the correction (Method 1) has been applied.  The eumelanin PL is heavily attenuated at all but the lowest concentration, but the expected linear relationship is fully recovered upon application of the correction.

\begin{figure}
	\begin{center}
		\includegraphics{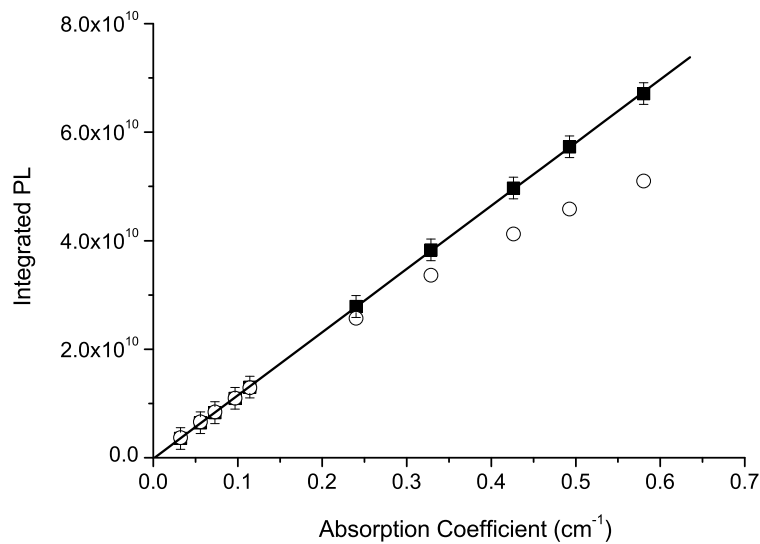}
	\end{center}
	\caption{Integrated PL emission vs. absorption coefficient at the excitation wavelength (490 nm) for ten fluorescein solutions ($1.2\times10-4\%$ to $5\times10^{-6}\%$ by weight): uncorrected data (open circles) and corrected with Method 1 (solid squares with linear regression). Errors were calculated according to equation \ref{eq:generaluncert}.}
	\label{fig:FluoQE}
\end{figure}

\begin{figure}
	\begin{center}
		\includegraphics{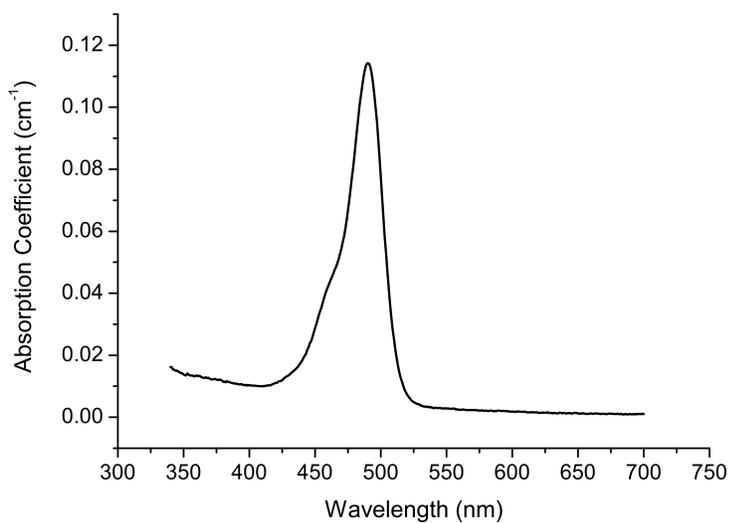}
	\end{center}
	\caption{Absorption spectrum of fluorescein solution (concentration by weight: $2\times10^{-5}\%$).}
	\label{fig:Fluoabs}
\end{figure}

As a standard fluorophore, fluorescein was used as a further test of the correction method (refer to Figure \ref{fig:FluoQE}).  Due to the very narrow absorbance range of fluorescein, unlike eumelanin, inner filter effects are far more significant than re-absorption.  Also unlike eumelanin, the very high quantum yield of fluorescein means that very low concentrations give measureable emission.  Five concentrations were used in the low concentration limit \cite{Horiba02,Williams83}, and five at higher concentrations as a test of the method.  The emission was attenuated as expected at these higher concentrations, but the expected linear relationship was fully recovered upon application of the correction (refer to Figure \ref{fig:FluoQE}).  Note that since fluorescein has a very sharp absorption peak (Figure \ref{fig:Fluoabs}), exciting at this peak (as we did here) will make inner filter effects far more significant than re-absorption ($\alpha_1d_1 \gg \alpha_2d_2$).

These measurements indicate that this correction method is valid, both for correction of re-absorption, and inner filter effects (or any combination of the two).


\subsection{Method 2:  Raman Peak Attenuation}
\begin{figure}
	\begin{center}
		\includegraphics{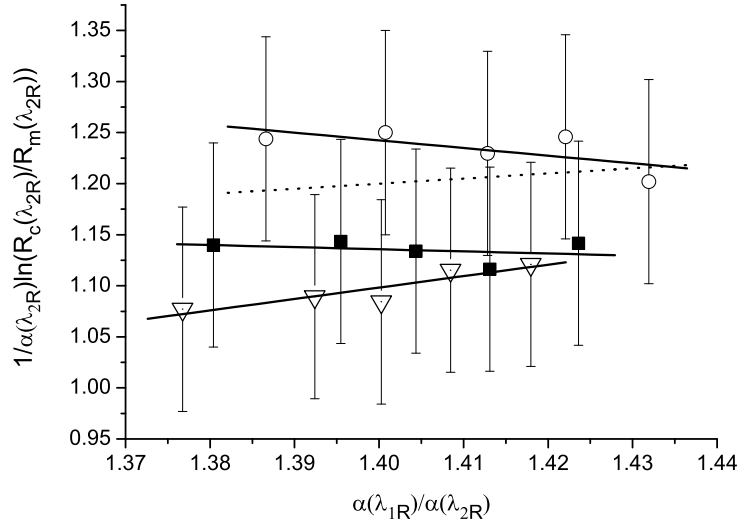}
	\end{center}
	\caption{Plot to determine $d_1$ and $d_2$, as described in the text.  Three solutions of different concentrations were measured:  0.001$\%$ (open circles), 0.0025$\%$ (filled squares) and 0.005$\%$ (open triangles).  The dotted line indicates the predicted trend, calculated from the known system geometry.  Errors were calculated according to equation \ref{eq:generaluncert} as described in the text.}
	\label{fig:d1d2plotALL}
\end{figure}

Figure \ref{fig:d1d2plotALL} shows the plot to determine $d_1$ (as the gradient) and $d_2$ (as the intercept) according to equation \ref{eq:Meth2}.  The three different solutions (of different concentrations) were measured in the same system, and so should all give identical values for $d_1$ and $d_2$ (and hence fall on the same line).  Figure \ref{fig:d1d2plotALL} shows that this is clearly not the case; although each data set appears to form a straight line, they are all significantly different, and none of the data sets gives close to the predicted values for $d_1$ and $d_2$ (as indicated by the dotted line).  Two of the data sets actually have negative gradients, and hence give negative values for $d_1$.

The uncertainty in each data point in Figure \ref{fig:d1d2plotALL} is very large, indicating that the linear trend in each data set is due to a systematic error in the fitting routine.  Hence we conclude that although this method of correction is theoretically correct, in practice there is too much freedom in the fitting process to determine $R_c$ and $R_m$ (and hence $d_1$ and $d_2$) accurately.  These parameters may be better determined by direct measurement (Method 1).  It is possible that this method may be viable for a sample with a larger Stokes shift; the Raman peak would then be separated further from the sample photoluminescence, making the fitting process more accurate.


\subsection{Method 3: Raman Peak Attenuation Approximation}
\begin{figure}
	\begin{center}
		\includegraphics{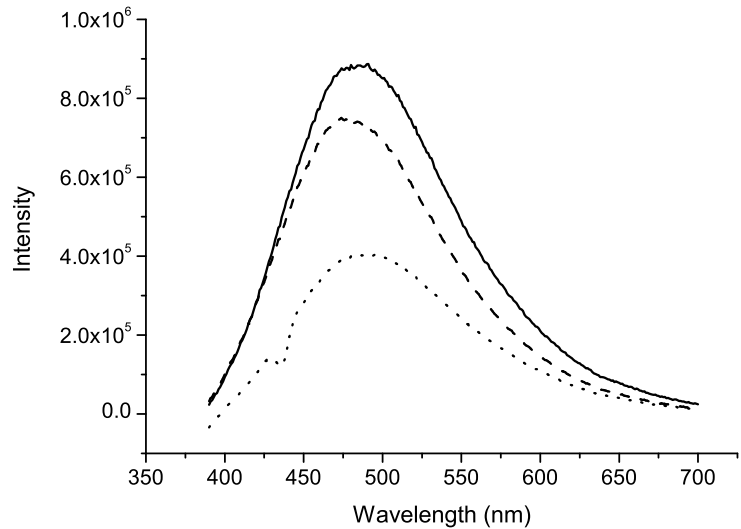}
	\end{center}
	\caption{PL of 0.0025$\%$ eumelanin solution (380nm excitation).  Solid:  corrected using Method 1, dashed: corrected using Method 3, dotted: no correction applied (background subtracted giving the artifact at 436 nm).}
	\label{fig:Meth1vs30025}
\end{figure}

Figure \ref{fig:Meth1vs30025} shows the PL of the 0.0025$\%$ eumelanin solution (excited at 380nm) corrected for re-absorption and inner filter effects with Methods 1 and 3, as well as the uncorrected data (with background subtracted).  Note that since the Raman scattering peak has been attenuated, subtraction of the background without prior correction leaves the artifact observed at 436nm.  The two methods agree well close to the Raman peak, but further away the approximation used in Method 3 is less accurate (as would be expected).  

\begin{figure}
	\begin{center}
		\includegraphics{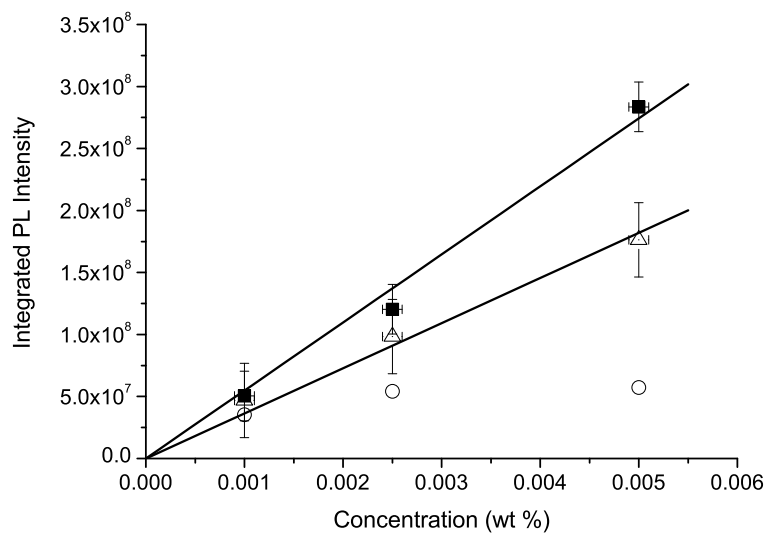}
	\end{center}
	\caption{Integrated PL of eumelanin solutions (380nm excitation) corrected with Method 1 (closed squares), Method 3 (open triangles) and uncorrected with background subtracted (open circles).   Linear regressions for each data set (shown) were constrained to pass through the origin.  Although at high concentrations Method 3 does not agree with Method 1 to within the uncertainty, it still provides a significant improvement over the uncorrected data.  Errors were calculated according to equations \ref{eq:generaluncert} and \ref{eq:uncert3PL}.}
	\label{fig:Methods1vs3}
\end{figure}

In Figure \ref{fig:Methods1vs3} we see the integrated PL against concentration for the two correction methods.  Both recover the expected linear relationship, and they agree at lower concentrations.  At high concentrations they deviate more significantly, indicating that if Method 3 must be used, extremely high concentrations should be avoided.  Method 3 still offers a vast improvement over the uncorrected data, however, as evidenced by Figure \ref{fig:Meth1vs30025}.

The effectiveness of Method 3 could not be investigated with the fluorescein solutions since the Raman scattering peak was of much lower intensity than the dye photoluminescence and hence not visible.  Note that this method could be applied to any emission or scattering feature that is detectable in both the background and sample spectra (not only the Raman peak).

\subsection{Correction of Photoluminescence Excitation Data}
\begin{figure}
	\begin{center}
		\includegraphics{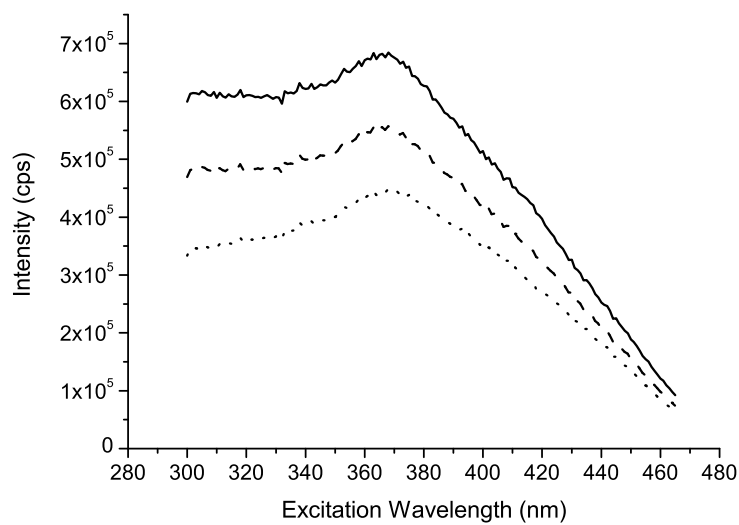}
	\end{center}
	\caption{Photoluminescence Excitation of eumelanin solution (0.001$\%$ by weight).  Uncorrected, with background subtracted (dotted), corrected with Method 1 (solid) and corrected with Method 3 (dashed).  The Raman peak height ratio for Method 3 was calculated from a PL spectrum excited at 380nm in the same geometry (rectangular cuvette with inner filter effects dominating).}
	\label{fig:PLEMethods1vs3}
\end{figure}

These correction methods can be equally applied to photoluminescence excitation (PLE) data.  For many materials, including melanins, accurate PLE data is critical for clear understanding of the emission behaviour.  Figure \ref{fig:PLEMethods1vs3} shows photoluminescence excitation data for a eumelanin solution.  The uncorrected data (with background subtracted) shows significant change when correction Method 1 is applied.  Method 3 does not agree closely, but still provides a significant improvement over the uncorrected data.  This is the first publication of PLE data for melanin, and will be expanded upon in future publications.


\section{Conclusion}
Two viable methods to correct for re-absorption and inner filter effects in emission and excitation spectra have been proposed and validated:
\begin{enumerate}
\item Direct Measurement (Method 1):  This method is the most accurate and easy to apply of those proposed.  Upon application of this correction all of the expected behaviours are recovered.  The geometry of the system must be well defined.
\item Raman Peak Attenuation Approximation (Method 3):  This method can be used even if the sample geometry is completely unknown (such as in thin films) as long as the Raman scattering peak is visible.  It is not as accurate as the first method, but gives a reasonable approximation, especially close to the Raman peak.
\end{enumerate}

Method 2 (exact correction via the Raman peak attenuation) failed in this case for technical reasons, but may be possible for situations where the Raman peak and fluorophore emission are better separated.

With these methods it is possible to recover the emission and excitation spectra of highly absorbing samples (such as melanin), which can be strongly distorted by re-absorption and inner filter effects.  These methods also make it possible to obtain far more accurate estimates of quantities such as the quantum yield in samples where high absorbance is unavoidable.  These methods have made it possible to obtain quantitative photoluminescence and photoluminescence excitation data for melanin, and most significantly the first accurate measurement of the quantum yield of melanin  \cite{Meredith04}.  This is a critical step towards understanding the energy dissipation pathways of this extremely important biological molecule.  

\section{Acknowledgments}
This work has been supported in part by the Australian Research Council, the UQ Centre for Biophotonics and Laser Science, and the University of Queensland (RIF scheme). Our thanks go to Paul Cochrane for writing the multiple Gaussian fitting routine, to Mark Fernee for his insight, and to Peter Riesz for his assistance with graphics.  



\end{document}